# CICV5G: A 5G Communication Delay Dataset for PnC in Cloud-based Intelligent Connected Vehicles

Xinrui Zhang, Peizhi Zhang, Junpeng Huang, Haojie Feng, Yining Ma, Feng Shen and Lu Xiong

*Abstract*—Cloud-based intelligent connected vehicles (CICVs) leverage cloud computing and vehicle-to-everything (V2X) to enable efficient information exchange and cooperative control. However, communication delay is a critical factor in vehicle-cloud interactions, potentially deteriorating the planning and control (PnC) performance of CICVs. To explore whether the new generation of communication technology, 5G, can support the PnC of CICVs, we present CICV5G, a publicly available 5G communication delay dataset for the PnC of CICVs. This dataset offers real-time delay variations across diverse traffic environments, velocity, data transmission frequencies, and network conditions. It contains over 300,000 records, with each record consists of the network performance indicators (e.g., cell ID, reference signal received power, and signal-to-noise ratio) and PnC related data (e.g., position). Based on the CICV5G, we compare the performance of CICVs with that of autonomous vehicles and examine how delay impacts the PnC of CICVs. The object of this dataset is to support research in developing more accurate communication models and to provide a valuable reference for scheme development and network deployment for CICVs. To ensure that the research community can benefit from this work, our dataset and accompanying code are made publicly available.

## I. INTRODUCTION

### A. Motivation

Recent advances in cloud computing and vehicle-to-everything (V2X) technology, cloud-based intelligent and connected vehicles (CICVs) are now at the forefront of the intelligent transportation system [1], [2]. The CICVs consist of the vehicle platform, communication network, and the cloud platform. The cloud platform collects information on the surrounding environment, vehicle status, and driving intentions through communication network. Leveraging the cloud platform's powerful computing capabilities, it performs PnC for multiple vehicles and sends the optimal trajectories commands or chassis control commands to vehicles [3-5]. It can significantly reduce the computational burden on vehicles and lower onboard system costs. Over the past decade, numerous scholars have proposed various methods for the PnC of CICVs. Hasan Esen *et al.* [6] successfully designs a cloud-based throttle control architecture. Simulation tests demonstrated the potential of CICVs. A cloud control scheme at intersections is proposed in [7], demonstrating its effectiveness in preventing collisions and alleviating traffic congestion. Li *et al.* introduce an integrated vehicle-road-cloud control system, further advancing the concept [8]. Numerous researchers have demonstrated the potential advantages of CICVs through simulation. However, they also point out that network communication delay poses a significant challenge to the practical application of CICVs.

Several strategies have been proposed for the PnC of CICVs under network delay conditions [9-17]. Among these methods, they construct communication delay models through reasonable assumptions or statistical modeling. Pan *et al.* [14] propose a cloud control method that accounts for communication quantization and stochastic delay, providing a theoretical approach for the safety control of CICVs. However, their experimental results are based on bounded random delay. Reference [15] proposes a consensus-based motion control algorithm to mitigate the impact of delay on control. However, they set the communication delay as a normal distribution with a mean of 40ms and a variance of 0.0259 in their simulation tests. Fang *et al.* [16] study the impact of communication delay on ramp merging cooperation and model delay as a normal distribution, proposing a delay compensation strategy. Nevertheless, due to the lack of real delay data on PnC, existing researches simplifies network communication delay as bounded random variable or assumes a normal distribution. Such simplified modeling methods struggle to reflect the variations in delay under different influencing factors. The fundamental issue lies in the lack of real-world delay data specific to the PnC.

### B. Literature Review of Communication Delay for PnC

Communication delay for PnC has been widely studied. On the one hand, some researchers have explored the performance of vehicle-to-network-to-vehicle (V2N2V) system in real-world environments, particularly in typical driving scenarios. The ping tool is commonly used for delay testing. References [18] and [19] employ it to assess 5G V2N2V communication delay. However, since the real data transmission protocols differ from the internet control message protocol (ICMP) used by the ping tool, this may result in an incomplete representation of the actual transmission delay in driving scenarios. Unlike the use of communication tools for testing, many researchers in the transportation field have measured round-trip time to calculate communication delay under real driving conditions. Wang *et al.* [20] test the variations in communication delay for V2X under different conditions, but the results lack relevant network performance metrics, overlooking the impact of network quality, a key communication indicator on communication delay. In recent studies, both closed-track testing [21-23] and on-road testing [24], [25] are employed. However, these tests focused primarily on the magnitude of communication delay and statistical results, without providing actual delay data. As a result, they lack a thorough investigation of the key factors and patterns that cause variations in communication delay, making it difficult to support the complex delay modeling requirements in PnC research.

On the other hand, communications engineering researchers have conducted studies on 5G V2N2V communication delay through theoretical analysis, simulation, and real-world measurements using professional tools. Based on the principles of communications, researchers have conducted delay modeling for various network components, including the radio access network [26], transmission network [27], and core network [28]. The literatures [29-32] have also addressed the construction of V2N2V network delay models. The primary aim of these studies is to enhance communication network design and optimize performance. Literature [33] conducts delay simulation testing through their open-source simulator. Furthermore, communications engineering researchers have also examined 5G communication delay in real-world scenarios. Raca *et al.* [34] propose a dataset that includes 5G network metrics related to vehicular entertainment and file downloads. The 5G-MOBIX project provides real-world test cases for 5G vehicular communications, employing professional network testing equipment for delay measurements. Their testing, however, is conducted at a frequency of 10Hz, which is not sufficient for the requirements of PnC [35]. Generally, these studies emphasize the transmission performance of 5G network and utilize straightforward, non-standard testing methods (e.g., simulation, and inadequate data size and frequency settings for PnC requirements). Consequently, these methods often struggle to meet the actual data interaction needs of CICVs and lack of quantitative analysis results regarding the real movement performance of vehicles under delay environments in PnC, which leads to an overestimation or underestimation of the impact of delay on the practical application of CICVs. In short, for the design of CICVs, communication delay data related to PnC is crucial. Developing such a dataset requires not only a real 5G network environment and relevant performance metrics but also data from actual driving scenarios, which is currently lacking.

*C. Contributions*

Motivated by this challenging situation, this paper proposes a dataset for 5G communication delay, specifically designed for the PnC of CICVs. The dataset focuses on the statistical characteristics of 5G communication delay in real driving environments, examining variations under different traffic conditions, driving velocity, data transmission frequency, and network signal strength. We establish a 5G delay testbed at the intelligent connected vehicle evaluation base at Tongji University and conduct extensive field tests. Through these tests, over 300,000 records are collected to build our dataset, CICV5G. The dataset includes not only communication delay and channel conditions (such as reference signal received power and signal-to-noise ratio) but also vehicle poses (i.e., vehicle coordinates, velocity). Based on CICV5G, we conduct a comparative analysis of CICVs and autonomous vehicles (AVs) performance in typical scenarios and explore the impact of communication delay on the PnC. To the best of our knowledge, this is the first publicly available dataset of 5G communication delay specifically for PnC of CICVs. The main contributions of this paper can be summarized as follows:

- A publicly available dataset for 5G V2N2V communication delay, specifically designed for the PnC of CICVs, is presented. This dataset addresses the lack of real data on research and provides essential support for network modeling.
- Several quantitative analyses of the PnC performance are obtained by on-track experiments. Testing results indicate that 5G communication can support the PnC of CICVs, achieving performance levels comparable to those of AVs under normal delay condition. Whereas, under a considerable delay, the time of delay occurrence has a greater impact on control errors. This analysis can be used to provide references for strategy and controller design.
- A modular and rapidly deployable 5G communication delay testbed for CICVs is conducted. It incorporates automated testing tools and procedures, making a positive contribution to standardized testing.

*D. Paper Organization*

The remaining sections of this paper are structured as follows. Section II presents an introduction to the attribution of our dataset. An overview production of the dataset is proposed in Section III. Section IV explores the impact of communication delay on the PnC of CICVs. Conclusions are drawn in Section V.

II. DATASET ATTRIBUTES DESCRIPTION

To meet the communication delay testing requirements, and based on industry standards set by the China Highway and Transportation Society [36], the dataset attributes have been designed as follows. The dataset primarily comprises PnC-related data and typical 5G network indicators. The former is gathered from the real-time inputs from the global navigation satellite system (GNSS) and vehicle chassis, while the latter is obtained using customer premises equipment (CPE) via the transmission control protocol (TCP) transmission protocol. To meet lightweight transmission requirements, all data is serialized using the protobuf protocol.

The attributes primarily include timestamps, position, heading angle, and driving velocity:
1) **Timestamp**
   Records the time when the vehicle sends data to the cloud (pub-time) and the time when it receives data from the cloud (sub-time). These two timestamps are recorded separately to accurately measure communication delay. To mitigate the effects of time desynchronization, round-trip time (RTT) is adopted.

2) **Position**
   Provides real-time vehicle position information, recorded in universal transverse mercator (UTM) format, with a recording accuracy of 0.01 meters.
3) **Heading**
   Includes the vehicle's heading angle information, recorded in the northeast up coordinate system using radians. The angle ranges from $[-\pi, \pi]$.
4) **Velocity**
   Records the vehicle's real-time velocity in meters per second (m/s).

Network performance indicators primarily include Cell ID, reference signal received power (RSRP), and signal-to-noise ratio (SINR):

1) **Cell ID**
   Identifies the 5G service cell in which the vehicle is located.
2) **RSRP**
   Measures the power level of the reference signal received at a specific location, which is commonly used to assess signal coverage and quality.
3) **SINR**
   Indicates the signal-to-noise ratio.

These metrics enable comprehensive analysis of the PnC of CICVs and such real-world test data are currently unavailable.

To meet the research needs, we provide a comprehensive and large-scale real-vehicle test data. The data collection ensures the inclusion of various traffic scenarios such as urban, rural, and highway, as well as varying 5G network conditions including private network and public network.

III. PRODUCTION DATASET OVERVIEW

This section provides details on the production of CICV5G used for the aforementioned protocol and testing. The data are collected using the testbed at Tongji University intelligent connected vehicle evaluation base.

*A. Testbed Design*

Firstly, we establish a standard testbed at the intelligent connected vehicle evaluation base of Tongji University. The 5G network comprehensively covers the three major testing zones: an urban test zone, an arterial road test zone supporting a maximum velocity of 80 km/h, and a rural and off-road test zone, as shown in Fig. 1(a). Fig. 1(b) illustrates the components and interaction flow of 5G delay tests for CICVs. The system consists three subsystems: the cloud platform, 5G communication network, and the vehicle platform.

The cloud contains cloud-based planning and control computation platform and the message queuing telemetry transport (MQTT) server. Data transmission uses the MQTT protocol, enabling information exchange through predefined publish/subscribe topics.

5G communication network is provided by the mobile network operators, with the network scheme including a 5G private network base station as well as public base stations. The 5G private network operates in the frequency band of 949.5-959.5 MHz, while the public network operates in the frequency band of 3400-3500 MHz. The 5G communication network consists of base stations and the core network, with the private network utilizing slicing technology to create an independent network for the test site.

The vehicle platform primarily handles trajectory tracking, including a data collection module, an onboard computing unit, and a 5G CPE. The vehicle encodes its real-time status information, including position, velocity, and heading angle, into Vsmdata and uploads it to the cloud.

The vehicle-cloud interaction process operates as follows: the vehicle transmits real-time status information (Vsmdata) to the 5G base station, which then forwards the data to the cloud via the network. The data is sent to the MQTT server, and after parsing and verification on the cloud, it is delivered to the computation unit. The planning results are subsequently encoded and sent back to the vehicle for tracking. This V2N2V communication architecture, where vehicles send status information to the cloud and receive trajectories in return, is also proposed with broad consensus [8]. Given this established framework, this paper focuses on examining the impact of 5G V2N2V delay on the PnC within this communication architecture.

Based on this testbed, a standardized testing procedure has been proposed, as shown in Fig. 1(c). The process commences by loading parameters to initiate the thread, followed by the creation of an MQTT client and verification of the network connection. Upon successful connection, the system encodes and transmits Vsmdata while monitoring the topic to receive CloudtoV data.

*B. Dataset Collection*

To ensure a diverse and comprehensive dataset, the experimental setup employs several statistical designs. We utilize a three-way factorial design in which the response is the communication delay with three independent variables or factors: scenario, 5G model, and velocity. Given the diverse requirements of autonomous driving scenarios, there may be correlations between different scenario factors, so we cannot treat the scenario factors as random. Therefore, a split-split-plot design (SSPD) is employed [37]. The Split-Split-Plot design is an experimental design method used to handle multiple factors, suitable for complex experimental setups. In this design, factors are divided into different levels, with each level of factors being experimented under every level of the preceding factors. This design can handle interaction effects between factors and effectively utilize resources.

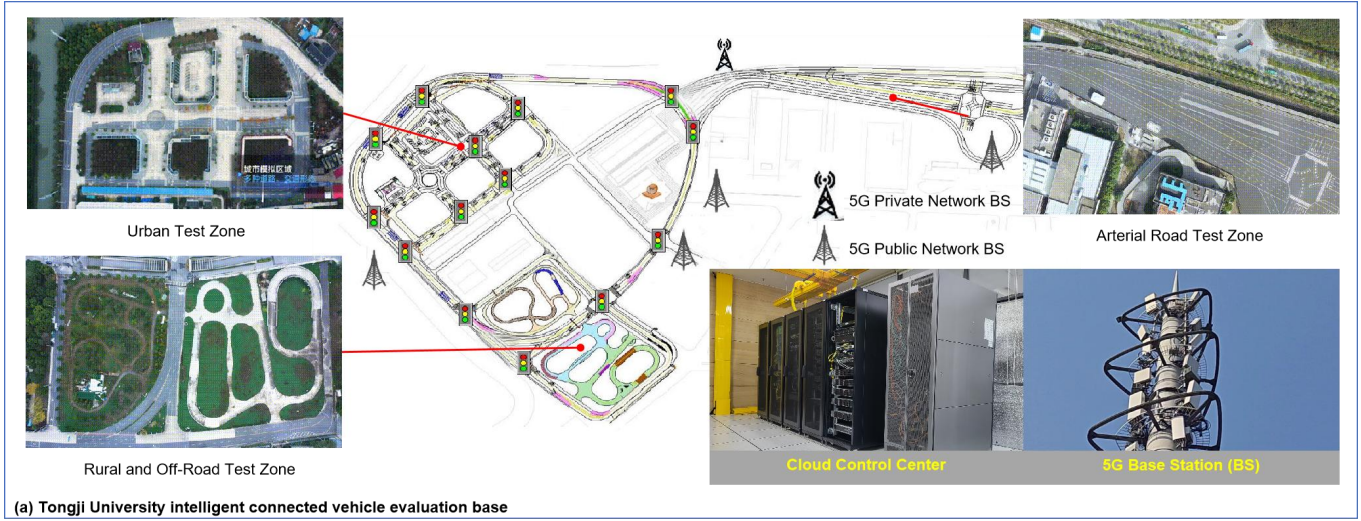
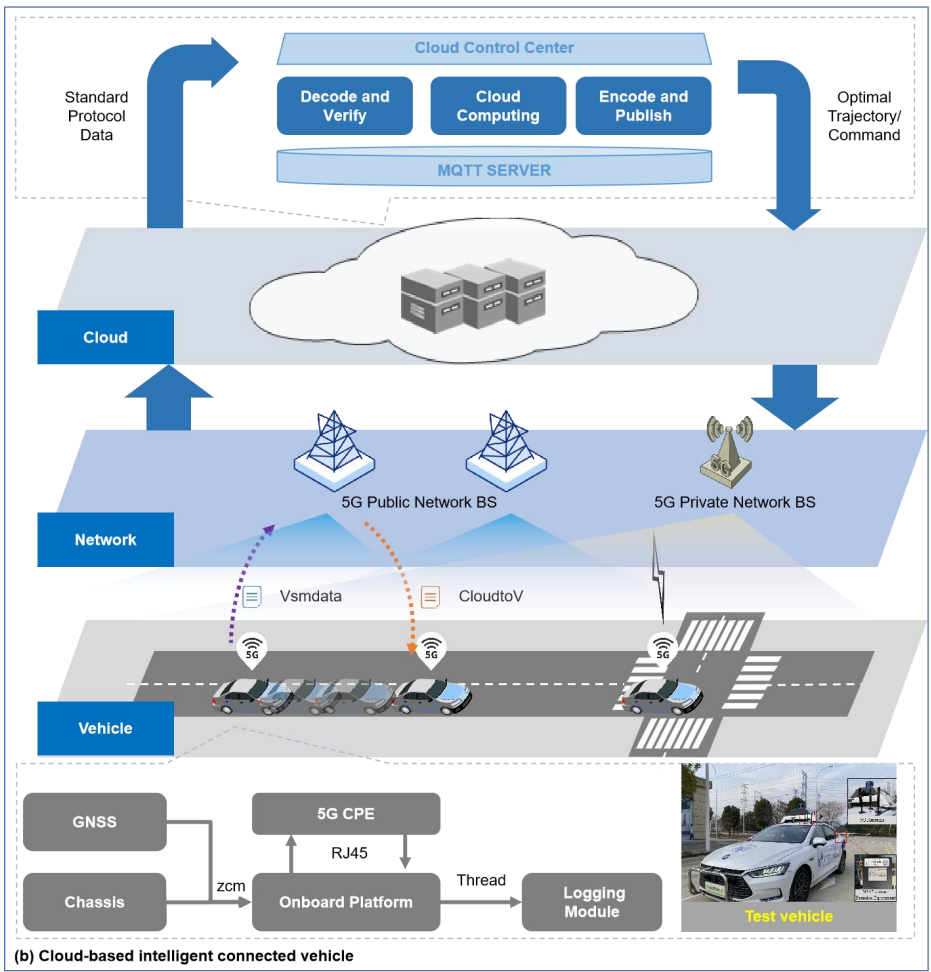
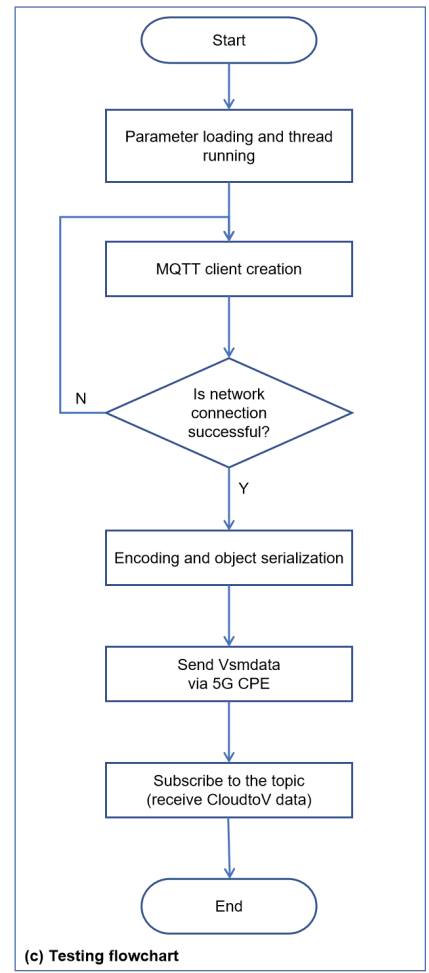

**Fig. 1.** Overview of Tongji University 5G communication delay testbed for CICVs.

### C. Exploration to the Statistical Characteristics of 5G Communication Delay

We test the 5G V2N2V communication delay in our testbed and select statistical indicators, including minimum value (Min), maximum value (Max), average delay (Avg), standard deviation (Std), 90th percentile (90th), and the proportion of latencies greater than 100 ms. The results based on the SSPD we designed are shown in Table I. It can be observed that in the regions with better network quality, such as the urban and arterial road test zones, the 5G V2N2V communication delay ranges from a minimum of 11 ms to a maximum of 694 ms. The average delay is approximately 18 ms, and the 90th percentile does not exceed 25 ms, indicating that the overall communication delay is relatively small. In terms of stability, the standard deviation of the 5G private network is smaller than that of the public network under the same testing environment and driving conditions, resulting in more stable communication delay. Notably, in the arterial road test zone, the proportion of delay exceeding 100 ms is lower for the private network compared to the public network.

From the comparison results in the rural and off-road test zone, where private network coverage is poor, it is evident that network signal quality has a significant impact on communication delay, leading to intolerably high delay and occasional disconnection issues, which include reconnection time in the delay measurements. Therefore, we report only the minimum delay values and the proportion of delay greater than 100 ms. It is apparent that unusually high delay can occur under certain specific conditions, which, although rare, do exist. To facilitate further research by other scholars on strategies and methods for handling unusually high delay conditions, we collect a substantial amount of data with weak signal quality but without disconnections in our testbed and construct a subset specifically for unusually high delay data.

Based on the test results, we observe that in zones with better signal quality, the proportion of abnormal delay is lower. Therefore, we perform data cleaning on the raw data and select four typical statistical models with the closest fitting performance: Gamma, Norm, Nakagami, and Rayleigh. We then use the residual sum of squares (RSS) and the akaike information criterion (AIC) as evaluation metrics. Detailed evaluations of the statistical results under various network modes and driving velocity conditions, based on these indicators, are provided in Appendix B. Based on the data under different operating conditions, we conclude that the Gamma distribution is the most suitable statistical model for 5G V2N2V communication delay in our tests, whether in private or public networks. The probability density function can be obtained using the following formula:

$$f(x) = \frac{b^a x^{a-1} e^{-bx}}{\Gamma(a)} \tag{1}$$

where $a$ and $b$ represent the shape and scale parameters of Gamma distribution, respectively.

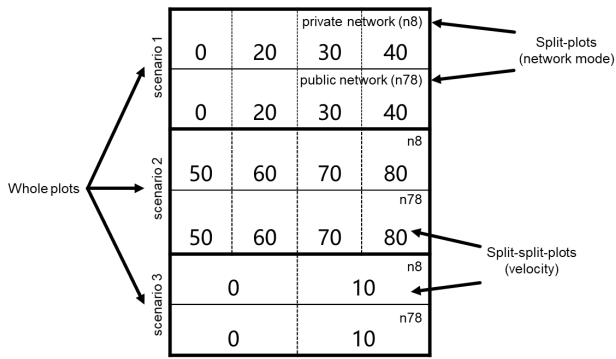

**Fig. 2.** Experimental design.

*D. Analysis of Factors Impacting 5G V2N2V Communication Delay*

**1) Velocity**

In real-world traffic environments, vehicles moving at high velocity frequently change the communication network topology, resulting in significant Doppler shift. This effect markedly degrades the network quality in 4G LTE-V2X [38]. It is essential to examine the effect of vehicle velocity on 5G V2N2V delay. To this end, we conduct empirical tests in the urban test zone, assessing typical urban driving velocity from 0 to 40 km/h, as well as in the arterial road test zone with high-velocity ranges from 50 to 80 km/h. These tests aim to explore the effects of velocity on 5G V2N2V communication delay.

The test results are illustrated in Fig. 3. Within the designed range of 0 to 80 km/h, not only in urban environments but also in arterial road scenarios, the delay value and deviation of 5G V2N2V remain nearly constant as driving velocity increases, with an approximate delay value of 18 ms. Furthermore, under the specified driving conditions, the impact of driving velocity on communication delay is not significant, regardless of whether in private or public networks. Through relevant researches, 5G adopts flexible subcarrier spacing and low-density parity check to mitigate the inter carrier interferences (ICIs) and provide stronger error correction capabilities. These improvements have significantly improved the communication performance of 5G compared to 4G in high-speed scenarios.

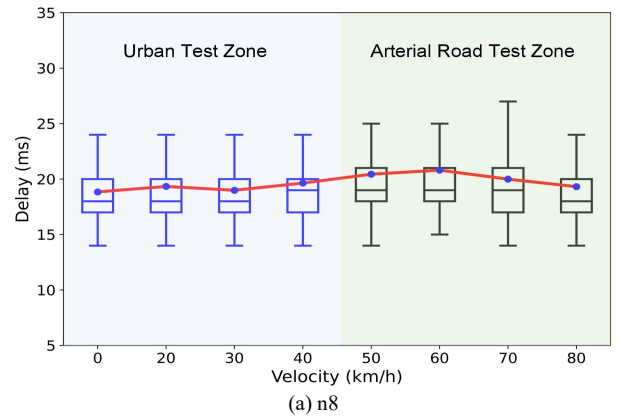

(a) n8

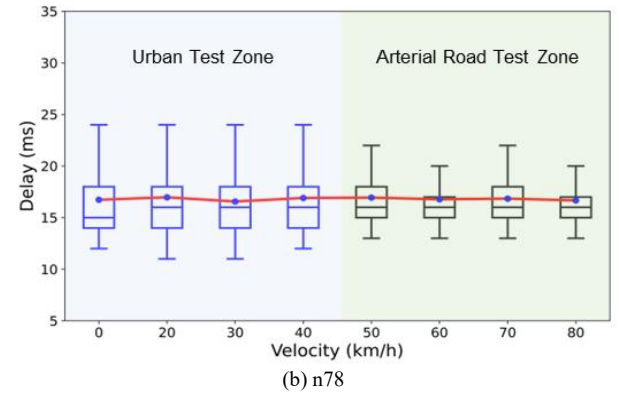

(b) n78

**Fig. 3.** Impact of velocity on 5G V2N2V delay.

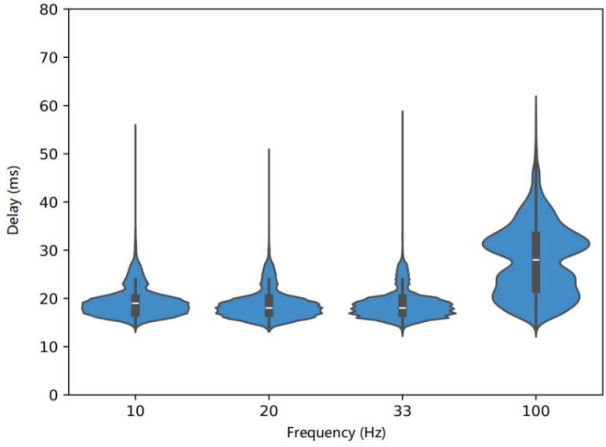

**Fig. 4.** Impact of data transmission frequency on 5G V2N2V delay.

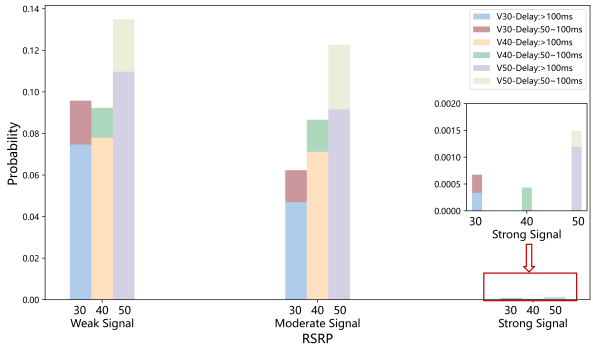

**Fig. 5.** Impact of RSRP on 5G V2N2V Delay.

#### 2) Data Transmission Frequency

In [39], various applications have distinct requirements for data transmission frequencies by 5G automotive association (5GAA). Therefore, we explore the impact of data transmission frequency on 5G V2N2V communication delay. We evaluate delay at frequencies of 10 Hz, 20 Hz, 33 Hz, and 100 Hz. To ensure consistency, all tests are conducted using the 5G private network at a speed of 30 km/h.

The impact of data transmission frequency on 5G communication delay is shown in Fig. 4. To analyze this impact, we focus on data with delay below 60 ms, and assess frequencies ranging from 10 to 100 Hz. For data transmission frequencies between 10 and 33 Hz, delay remain relatively stable, with an average of approximately 18.7 ms and a standard deviation of about 2.9 ms. However, when data transmission frequency reaches 100 Hz, delay significantly increases, with the average rising to 28 ms and delay stability deteriorating, with the standard worsening to 7.3 ms. This increase in delay is likely due to data congestion associated with higher data transmission frequencies. The arrival of data packets exceeds the network processing capacity, resulting in queuing delay, or the loss of preceding packets triggers the retransmission mechanism, causing retransmission delay.

#### 3) Network Signal Strength

To evaluate the impact of network signal strength on 5G V2N2V communication delay, we categorize RSRP into three levels: strong signal (> -85 dBm), moderate signal (-90 to -85 dBm), and weak signal (< -90 dBm). We then calculate the different proportions of communication delay exceeding 50 ms and 100 ms for each RSRP category.

The impact of network signal strength on 5G V2N2V communication delay is illustrated in Fig. 5. Under the speed of 30, 40, and 50 km/h, the proportion of high delay increases significantly as the RSRP transitions from strong to weak. This demonstrates a substantial effect of RSRP on communication delay. Specifically, in conditions of weak signal quality, the proportion of high delay can be up to 180 times greater compared to strong signal conditions.

*E. Hypothesis Testing on the Impact of Communication Delay on Vehicle Pose Deviation*

For the PnC of CICVs, the presence of communication delay introduces the deviations in vehicle state information, which in turn affects the performance of PnC. As illustrated in Fig. 1(b), the vehicle uploads its state information to the cloud at time $t_k$. The cloud uses this information to plan its trajectory, which the vehicle receives at time $t_{k+d}$. However, by this time the vehicle receives the planned trajectory, its state has already changed. Thus, the time-varying nature of communication delay manifests as the uncertainty in vehicle state. Based on the aforementioned test results, RSRP is a critical factor influencing communication delay, and driving velocity also significantly impacts on the PnC. Therefore, we test the deviation between the vehicle pose received by the cloud at a driving speed of 30 km/h and the actual vehicle pose.

The impact of communication delay on vehicle pose deviation is illustrated in Fig. 6. In Fig. 6(a), it can be observed that the RSRP generally decreases from the start point to the end point, with a noticeable phenomenon of abrupt changes. Fig. 6(b) depicts the variation in communication delay, revealing that worse signal strength correlates with a significant increase in delay. Particularly when signal strength deteriorates abruptly, leading to a maximum delay of up to 480 ms. Fig. 6(c) displays the deviation between the transmitted vehicle position and the actual vehicle position, with a maximum deviation of 4.02 m. Combined with Fig. 6(a-c), deteriorating RSRP increases the likelihood of high communication delay. There is a proportional relationship between vehicle position deviation and communication delay, specifically, the pose deviation is the product of the delay and vehicle velocity.

Therefore, analysis of variance is adopted to explore the impact of the RSRP and velocity on position deviation. We let the RSRP (R) have $r$ levels, denoted as $R_1, R_2, \ldots R_r$, and let the velocity (V) have v levels, denoted as $V_1, V_2, \ldots V_v$. For each combination of these levels, $c$ repeated experiments are conducted. The result of the $k$-th experiment under the combination $R_i \times V_j$ is denoted as $X_{ijk}$. Each experiment is repeated

TABLE I
THE TEST RESULTS OF 5G V2N2V COMMUNICATION DELAY

| Scenarios | Network | Velocity (km/h) | Min (ms) | Max (ms) | Avg (ms) | Std (ms) | 90th (ms) | $\geq$ 100ms (%) |
|---|---|---|---|---|---|---|---|---|
| Urban test zone | n8 | 0 | 14 | 274 | 18.84 | 6.93 | 23 | 0.12 |
| | | 20 | 14 | 343 | 19.33 | 10.22 | 23 | 0.23 |
| | | 30 | 14 | 270 | 18.98 | 7.48 | 23 | 0.13 |
| | | 40 | 14 | 269 | 19.64 | 9.95 | 24 | 0.25 |
| | n78 | 0 | 12 | 265 | 16.73 | 9.16 | 20 | 0.19 |
| | | 20 | 11 | 694 | 16.97 | 14.07 | 20 | 0.25 |
| | | 30 | 11 | 304 | 16.56 | 7.15 | 20 | 0.11 |
| | | 40 | 12 | 277 | 16.90 | 9.85 | 20 | 0.24 |
| Arterial road test zone | n8 | 50 | 14 | 288 | 20.44 | 8.71 | 25 | 0.17 |
| | | 60 | 15 | 287 | 20.80 | 13.49 | 24 | 0.47 |
| | | 70 | 14 | 316 | 19.99 | 10.36 | 24 | 0.23 |
| | | 80 | 14 | 287 | 19.31 | 10.70 | 22 | 0.30 |
| | n78 | 50 | 13 | 323 | 18.91 | 19.28 | 21 | 0.89 |
| | | 60 | 13 | 280 | 18.39 | 17.70 | 21 | 0.85 |
| | | 70 | 13 | 279 | 19.35 | 21.81 | 21 | 1.27 |
| | | 80 | 13 | 271 | 19.13 | 21.03 | 21 | 1.25 |
| Rural and off-road test zone | n8 | 0 | 14 | 1568 | 118.70 | 221.41 | 394 | 22.01 |
| | | 10 | 14 | Occurrence of disconnections | | | | 60.58 |
| | n78 | 0 | 11 | 261 | 16.57 | 11.98 | 19 | 0.37 |
| | | 10 | 11 | 286 | 17.18 | 15.00 | 20 | 0.58 |

multiple times, and the results are independent of each other. The model is established in equation 2-8:

$$X_{ijk} \sim N(\mu_{ij},\sigma^2), i=1,2,\ldots,r; j=1,2,\ldots,v; k=1,2,\ldots,c \quad (2)$$

$$\mu = \frac{1}{rv}\sum_{i=1}^{r}\sum_{j=1}^{v}\mu_{ij} \quad (3)$$

$$\bar{\mu}_{i\cdot} = \frac{1}{v}\sum_{j=1}^{v}\mu_{ij} \quad (4)$$

$$\bar{\mu}_{\cdot j} = \frac{1}{r}\sum_{i=1}^{r}\mu_{ij} \quad (5)$$

$$\alpha_i = \bar{\mu}_{i\cdot} - \mu \quad (6)$$

$$\beta_j = \bar{\mu}_{\cdot j} - \mu \quad (7)$$

$$\mu_{ij} = \mu + \alpha_i + \beta_j + \gamma_{ij} \quad (8)$$

where $\alpha_i$ represents the effect of RSRP at the $i$-th level, $\beta_j$ represents the effect of velocity at the $j$-th level, $\gamma_{ij}$ represents the interaction effect. The following hypotheses are proposed:

$$\begin{aligned} H_{01} &: \alpha_1 = \alpha_2 = \cdots = \alpha_r = 0 \\ H_{02} &: \beta_1 = \beta_2 = \cdots = \beta_v = 0 \\ H_{03} &: \gamma_{ij} = 0, i=1,2,\ldots,r; j=1,2,\ldots,v \end{aligned} \quad (9)$$

TABLE II
ANALYSIS OF VARIANCE

| Source | Sum of squared deviations | Degrees of freedom | Mean squared error | F-value |
|---|---|---|---|---|
| R | 0.0399 | 2 | 0.0195 | 5.4167 |
| V | 0.0136 | 2 | 0.0068 | 1.8889 |
| I | 0.0481 | 4 | 0.0120 | 3.3333 |

| Error | 0.0652 | 18 | 0.0036 |
| Total | 0.3828 | 26 | |

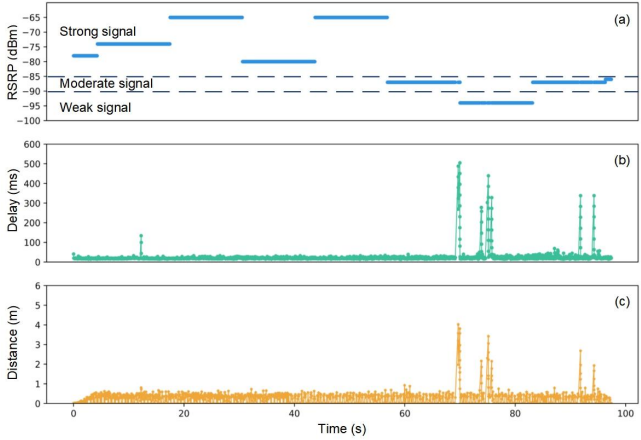

**Fig. 6.** Impact of communication delay on position deviation at a speed of 30 km/h.

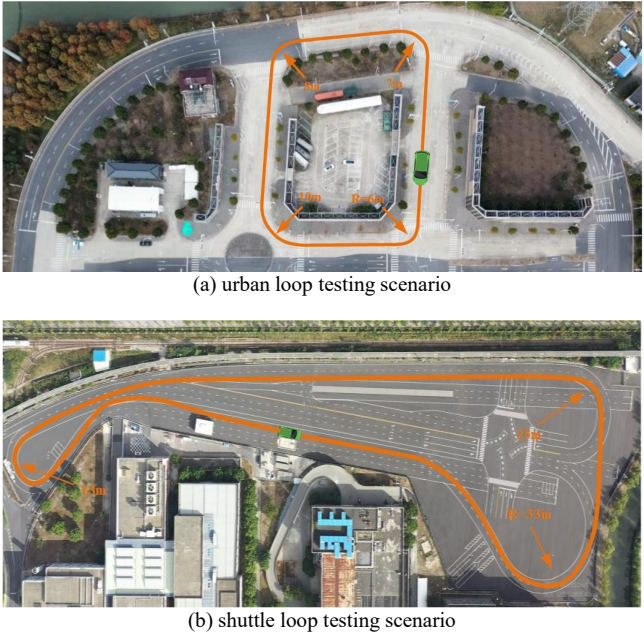

(a) urban loop testing scenario

(b) shuttle loop testing scenario

**Fig. 7.** Two testing scenarios.

Leveraging the experimental data, calculate the values of the statistical metrics, construct the analysis of variance table, and present the results as shown in Table II.

$$F_R > F_{0.05}(2,18) = 3.55$$
$$F_V < F_{0.05}(2,18) = 3.55 \quad (10)$$
$$F_I > F_{0.05}(4,18) = 2.73$$

Based on Eq. (10), it is concluded at a significance level of 0.05 that RSRP has a significant effect on vehicle pose deviation, while velocity does not. Additionally, the interaction between RSRP and velocity significantly impacts vehicle pose deviation.

## IV. EXPERIMENTAL EVALUATION

In order to explore the actual effect of communication delay on the PnC performance of CICVs, this section conduct testing focus on PnC under real driving conditions. Utilizing the dataset, we select typical conditions to conduct on-site real vehicle testing. Taking the PnC of AVs as the baseline, we compare the actual performance of CICVs with that of AVs, and explore the impact of 5G communication delay on the PnC of CICVs.

### A. Design of the PnC Strategy for CICVs and AVs

Firstly, we design two strategies: one for the PnC of CICVs and another for AVs. To ensure experimental consistency, both strategies utilize a lattice algorithm at the planning level, with longitudinal tracking control using a PID controller and lateral tracking control using the Stanley algorithm. The primary distinction is in the cloud control strategy, where the planning algorithm is deployed in the cloud. This method involves receiving vehicle status information, processing it through the cloud-based planning module, and sending the resulting trajectories back to the vehicle for tracking. In contrast, the AVs strategy processes all functions internally within the vehicle.

### B. Comparison of the Motion Performance between CICVs and AVs

In this experiment, the vehicle used, as shown in Fig. 1(b), is a modified BYD Qin-Pro electric vehicle. The tests are conducted within our testbed. Two scenarios are designed, as shown in Fig. 7, a) *urban loop* with repeated right turns. It consists of four curved sections connected by straight lines, ideal for testing the step response at urban test zone, and b) *shuttle loop*, which has two straight sections connected by turn-around sections with varying but continuous curvatures at arterial road test zone.

According to the testing flowchart shown in Fig. 1(c), the test results under urban loop conditions are shown in Fig. 8. The blue line represents the performance of the AV's PnC scheme, while the red line denotes the performance of the CICVs. Under this condition, as shown in Fig. 8(a), the route consists of four quarter circles with constant radii 6, 10, 8, and 7m. The curvature profile is step-shaped. Fig. 8(b) shows the comparison of lateral error between CICVs and AVs. The maximum lateral error $e_y$ for the PnC of AVs is 20.08 cm, compared to 20.75 cm for the CICVs. Fig. 8(c) shows the comparison of heading error between CICVs and AVs. The maximum heading error $e_\theta$ is 0.05 rad for AVs and 0.06 rad for CICVs. Fig. 8(d) shows the vehicle velocity. As a comparison, both PnC schemes can achieve stable control effects. The reason for this is that under normal delay, the PnC strategy itself has the ability to resist control errors and can approximately achieve control effect similar to that of AVs. Comparison of testing results between the two refer to the supplementary videos:
1) Video1_urbanloop_AV_PnC.mp4;
2) Video2_urbanloop_CICV_PnC.mp4.

Fig.9 illustrates the results of the shuttle loop scenario. In the testing process over a distance of nearly 1 km, the performances of the two PnC strategy are comparable. In terms of lateral error, the average errors for AV and CICV are 0.11 m and 0.16 m, respectively. Regarding heading error, the average errors are 0.04 rad for AV and 0.05 rad for CICV. Although the cloud-based approach is slightly less effective due to communication delay, both strategies maintain motion stability. One reason is that this scenario with continuous and smooth road curvatures is less sensitive to communication delay. Comparison of testing results between the two refer to the supplementary videos:
1) Video3_shuttleloop_AV_PnC.mp4;
2) Video4_shuttleloop_CICV_PnC.mp4.

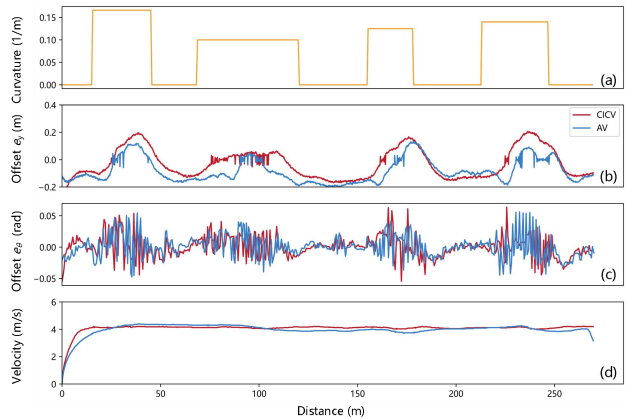

**Fig. 8.** Experimental results of the urban loop.

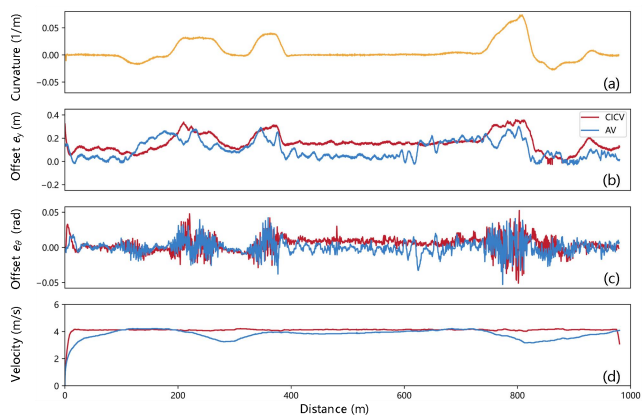

**Fig. 9.** Experimental results of the shuttle loop.

*C. Exploration of the Impact of 5G Communication Delay*

In order to explore the performance of CICVs' PnC under time-varying abnormal delay, we set three types of large delay by thread sleep, based on the delay results in Table I. These delays are: 100-200ms (delay1), 200-400ms (delay2), and 500-700ms (delay3), all with random large delay settings.

The results in Fig. 10 and 11 indicate that, despite the varying magnitudes of delay, vehicles relying on PnC strategies maintain a lateral error of no more than 0.07 m at the straight segments. The impact of abnormal large delay on the motion of CICVs is not significant. However, on curves with varying curvature, as the delay worsens, the lateral error of the vehicle increases continuously. When the communication delay is in the range of 500-700 ms, the maximum lateral error can reach up to 0.65 m. Comparing the results in Fig. 10 with those in Fig. 6 on the impact of communication delay on position deviation, we observe an interesting finding: under constant-speed driving conditions, larger communication delays lead to increased vehicle position errors, demonstrating a positive correlation that is dependent on driving speed. By contrast, under the influence of PnC strategies, the actual tracking error of the vehicle exhibits a nonlinear relationship with communication delay. In other words, the deterioration of communication delay impacts the motion control of CICVs, but the extent of this impact depends not only on the magnitude of the delay but also on the timing of its occurrence. For PnC strategies, within a certain range of delays, the vehicle system has the capability to ensure motion stability. As illustrated in Figure 11, it can be inferred that if the delay continues to increase, or if sporadic disconnections occur, vehicle tracking errors will rise, potentially leading to system instability. This stability is closely related to the PnC strategy, emphasizing the importance of utilizing more comprehensive and accurate delay data to construct a precise delay model.

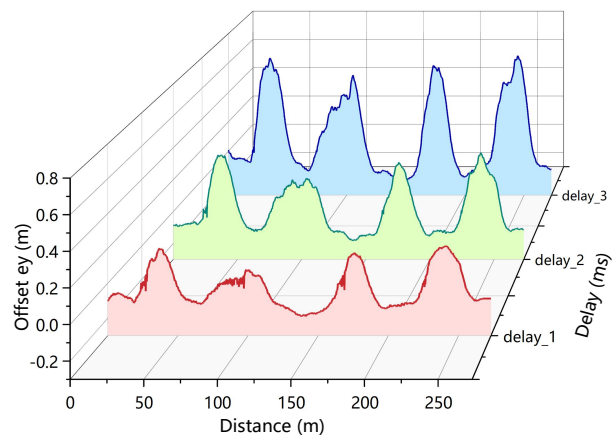

**Fig. 10.** Results of lateral offset under different time delay.

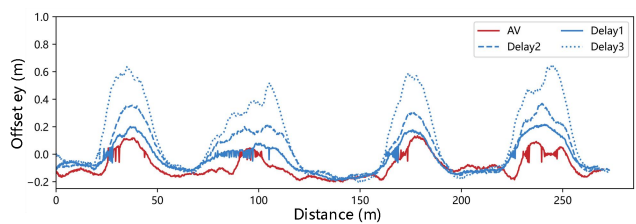

**Fig. 11.** Comparison of lateral offset with different time delay between CICVs and AVs.

V. CONCLUSION

This paper presents a 5G V2N2V communication delay dataset dedicated to the PnC of CICVs, namely CICV5G. It consists of vehicle pose information and network indicators related to 5G communication delay. To obtain accurate and comprehensive data, a 5G communication delay testbed has been developed at the intelligent connected vehicle evaluation base of Tongji University. The effects of driving velocity, data transmission frequency, and the reference signal received power on communication delay are explored through real-world vehicle testing. The impact of communication delay on vehicle motion performance is examined. In addition, we test the impact of 5G V2N2V delay on the PnC. The empirical results indicate that our dataset is suitable for exploring the impact of delay on the PnC of CICVs, as well as for designing PnC strategies and operational design domain (ODD) for CICVs. In summary, the proposed dataset addresses the lack of real data in the design of PnC strategies, provides a reference for establishing a standardized 5G V2N2V delay test, and makes a positive contribution to the practical implementation of CICVs.

## APPENDIX

### A. Probability Density Function Fitting Curve of 5G V2N2V Communication Delay Data Statistics in Different Scenarios

In each sub-graph, the horizontal axis represents the delay value (unit: ms) and the vertical axis represents the probability density. Lines with different color are different fitting probability density functions, including Normal, Gamma, Nakagami, and Rayleigh distribution. Fig 12-14 represent the results of the different test zones in our testbed.

### B. Evaluation of Statistical Models for Fitting 5G V2N2V Communication Delay

Table III-VI present the evaluation results of the different test zones in our testbed.

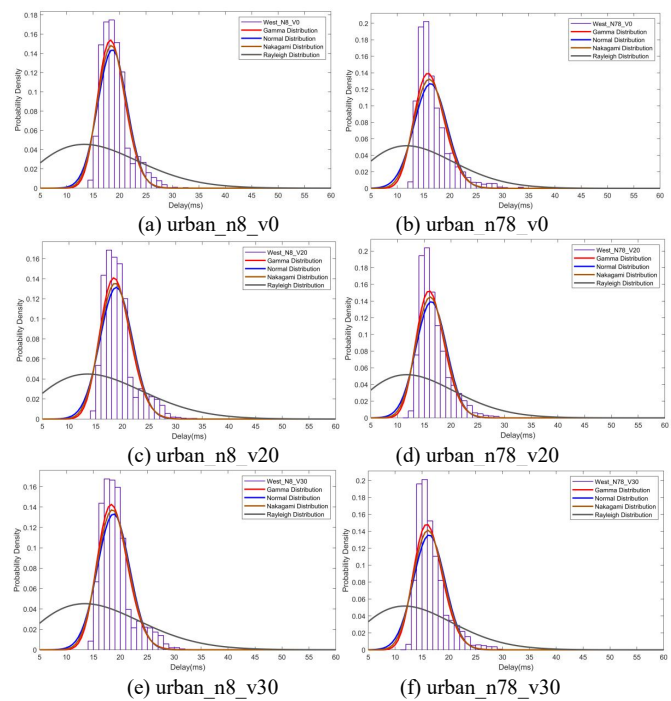

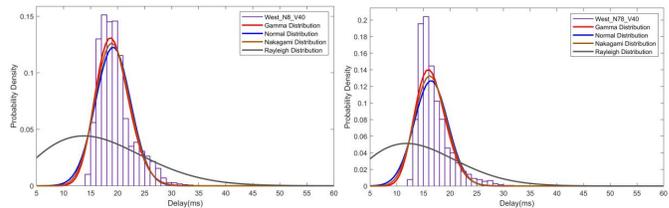

**Fig. 12.** Probability density function of 5G V2N2V in urban test zone.

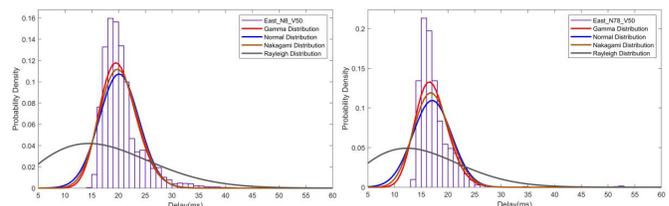

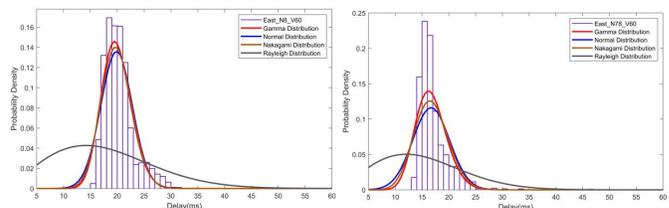

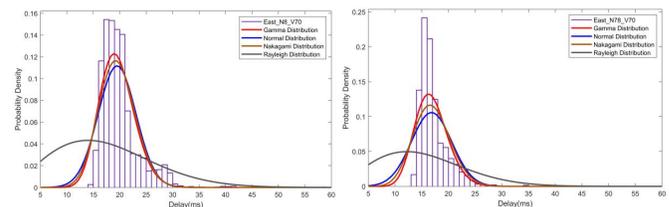

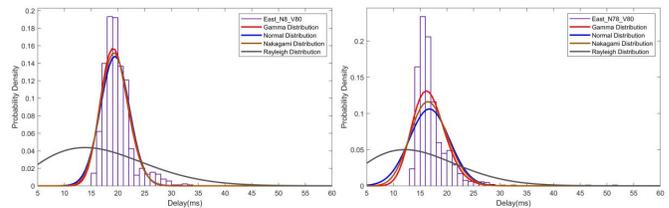

**Fig. 13.** Probability density function of 5G V2N2V in arterial road test zone.

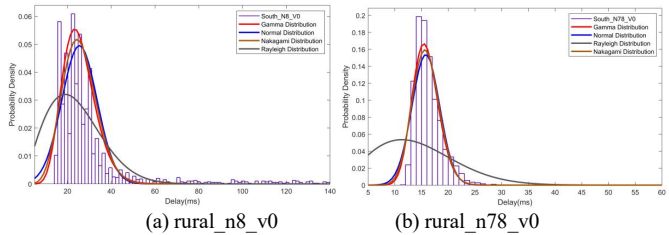

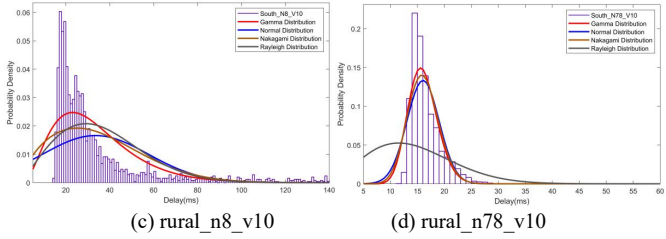

(c) rural_n8_v10  (d) rural_n78_v10

**Fig. 14.** Probability density function of 5G V2N2V in rural and off-road test zone.

TABLE III
EVALUATION RESULTS FOR 5G N8 IN URBAN TEST ZONE

| Velocity (km/h) | statistical model | evaluating indicator RSS | AIC |
|---|---|---|---|
| 0 | normal | 0.0141 | -10.0838 |
| 0 | nakagami | 0.0115 | -10.4892 |
| 0 | rayleigh | 0.0768 | -6.7866 |
| 0 | **gamma** | **0.0090** | **-10.9810** |
| 20 | normal | 0.0178 | -9.6119 |
| 20 | nakagami | 0.0149 | -9.9719 |
| 20 | rayleigh | 0.0718 | -6.9210 |
| 20 | **gamma** | **0.0120** | **-10.4083** |
| 30 | normal | 0.0170 | -9.7010 |
| 30 | nakagami | 0.0141 | -10.0774 |
| 30 | rayleigh | 0.0724 | -6.9054 |
| 30 | **gamma** | **0.0112** | **-10.5372** |
| 40 | normal | 0.0136 | -10.1561 |
| 40 | nakagami | 0.0111 | -10.5645 |
| 40 | rayleigh | 0.0589 | -7.3193 |
| 40 | **gamma** | **0.0087** | **-11.0557** |

TABLE IV
EVALUATION RESULTS FOR 5G N78 IN URBAN TEST ZONE

| Velocity (km/h) | Statistical model | evaluating indicator RSS | AIC |
|---|---|---|---|
| 0 | normal | 0.0265 | -8.8151 |
| 0 | nakagami | 0.0226 | -9.1353 |
| 0 | rayleigh | 0.0688 | -7.0080 |
| 0 | **gamma** | **0.0185** | **-9.5350** |
| 20 | normal | 0.0217 | -9.2173 |
| 20 | nakagami | 0.0182 | -9.5678 |
| 20 | rayleigh | 0.0720 | -6.9166 |
| 20 | **gamma** | **0.0147** | **-9.9984** |
| 30 | normal | 0.0235 | -9.0613 |
| 30 | nakagami | 0.0197 | -9.4070 |
| 30 | rayleigh | 0.0719 | -6.9184 |
| 30 | **gamma** | **0.0159** | **-9.8369** |
| 40 | normal | 0.0263 | -8.8345 |
| 40 | nakagami | 0.0223 | -9.1614 |
| 40 | rayleigh | 0.0700 | -6.9725 |
| 40 | **gamma** | **0.0182** | **-9.5733** |

TABLE V
EVALUATION RESULTS FOR 5G N8 IN ARTERIAL ROAD TEST ZONE

| Velocity (km/h) | statistical model | evaluating indicator RSS | AIC |
|---|---|---|---|
| 0 | normal | 0.0194 | -9.4365 |
| 0 | nakagami | 0.0162 | -9.8077 |
| 0 | rayleigh | 0.0620 | -7.2147 |
| 0 | **gamma** | **0.0127** | **-10.2825** |
| 20 | normal | 0.0146 | -10.0098 |
| 20 | nakagami | 0.0120 | -10.4065 |
| 20 | rayleigh | 0.0767 | -6.7908 |
| 20 | **gamma** | **0.0094** | **-10.8932** |
| 30 | normal | 0.0205 | -9.3342 |
| 30 | nakagami | 0.0170 | -9.6998 |
| 30 | rayleigh | 0.0645 | -7.1347 |
| 30 | **gamma** | **0.0135** | **-10.1644** |
| 40 | normal | 0.0150 | -9.9589 |
| 40 | nakagami | 0.0124 | -10.3374 |
| 40 | rayleigh | 0.0870 | -6.5376 |
| 40 | **gamma** | **0.0099** | **-10.7948** |

TABLE VI
EVALUATION RESULTS FOR 5G N78 IN ARTERIAL ROAD TEST ZONE

| Velocity (km/h) | statistical model | evaluating indicator RSS | AIC |
|---|---|---|---|
| 0 | normal | 0.0374 | -8.1296 |
| 0 | nakagami | 0.0314 | -8.4787 |
| 0 | rayleigh | 0.0819 | -6.6579 |
| 0 | **gamma** | **0.0244** | **-8.9814** |
| 20 | normal | 0.0479 | -7.6326 |
| 20 | nakagami | 0.0413 | -7.9311 |
| 20 | rayleigh | 0.0965 | -6.3310 |
| 20 | **gamma** | **0.0333** | **-8.3585** |
| 30 | normal | 0.0507 | -7.5187 |
| 30 | nakagami | 0.0433 | -7.8352 |
| 30 | rayleigh | 0.0950 | -6.3621 |
| 30 | **gamma** | **0.0340** | **-8.3192** |
| 40 | normal | 0.0501 | -7.5422 |
| 40 | nakagami | 0.0430 | -7.8514 |
| 40 | rayleigh | 0.0935 | -6.3943 |
| 40 | **gamma** | **0.0340** | **-8.3181** |

REFERENCES

[1] K. Li, J. Wang, and Y. Zheng, "Cooperative Formation of Autonomous Vehicles in Mixed Traffic Flow: Beyond Platooning," *IEEE Trans. Intell. Transp. Syst.*, vol. 23, no. 9, pp. 15951–15966, Sep. 2022, doi: 10.1109/TITS.2022.3146612.

[2] Z. Li, J. Hu, B. Leng, L. Xiong, and Z. Fu, "An Integrated of Decision Making and Motion Planning Framework for Enhanced Oscillation-Free Capability," *IEEE Trans. Intell. Transp. Syst.*, vol. 25, no. 6, pp. 5718–5732, Jun. 2024, doi: 10.1109/TITS.2023.3332655.

[3] J. A. Guerrero-ibanez, S. Zeadally, and J. Contreras-Castillo, "Integration challenges of intelligent transportation systems with

[3] connected vehicle, cloud computing, and internet of things technologies," *IEEE Wirel. Commun.*, vol. 22, no. 6, pp. 122–128, Dec. 2015, doi: 10.1109/MWC.2015.7368833.
[4] X. Zhang, L. Xiong, P. Zhang, B. Leng, and Y. Che, "Cloud Control with Communication Delay Prediction for Intelligent Connected Vehicles," in *Proc. IEEE Intell. Vehicles Symp. (IV)*, Jun. 2024, pp. 539–544. doi: 10.1109/IV55156.2024.10588607.
[5] J. Dong, Q. Xu, J. Wang, C. Yang, M. Cai, C. Chen, Y. Liu, J. Wang, and K. Li, "Mixed Cloud Control Testbed: Validating Vehicle-Road-Cloud Integration via Mixed Digital Twin," *IEEE Trans. Intell. Veh.*, vol. 8, no. 4, pp. 2723–2736, Apr. 2023.
[6] H. Esen, M. Adachi, D. Bernardini, A. Bemporad, D. Rost, and J. Knodel, "Control as a service (CaaS): cloud-based software architecture for automotive control applications," in *Proc. ACM SWEC*, Apr. 2015, pp. 13-15.
[7] M. A. Nazari, T. Charalambous, J. Sjoberg, and H. Wymeersch, "Remote control of automated vehicles over unreliable channels," in *Proc. 2018 IEEE Wireless Comm. Networking (WCNC)*, Apr. 2018, pp. 1–6.
[8] W. Chu, Q. Wuniri, X. Du, Q. Xiong, T. Huang, and K. Li, "Cloud Control System Architectures, Technologies and Applications on Intelligent and Connected Vehicles: a Review," *Chin. J. Mech. Eng.*, vol. 34, no. 1, Dec. 2021, doi: 10.1186/s10033-021-00638-4.
[9] T. Zeng, O. Semiari, W. Saad, and M. Bennis, "Joint Communication and Control for Wireless Autonomous Vehicular Platoon Systems," *IEEE Trans. Commun.*, vol. 67, no. 11, pp. 7907–7922, Nov. 2019, doi: 10.1109/TCOMM.2019.2931583.
[10] T. Zeng, O. Semiari, W. Saad, and M. Bennis, "Joint Communication and Control System Design for Connected and Autonomous Vehicle Navigation," in *ICC 2019 - 2019 IEEE International Conference on Communications (ICC)*, May 2019, pp. 1–6.
[11] M. P. Vitus, Z. Zhou, and C. J. Tomlin, "Stochastic Control With Uncertain Parameters via Chance Constrained Control," *IEEE Trans. Autom. CONTROL*, vol. 61, no. 10, 2016.
[12] B. Tian, "Communication delay compensation for string stability of CACC system using LSTM prediction," *Veh. Commun.*, 2021.
[13] J. Pan, Q. Xu, K. Li, and J. Wang, "A vehicle cloud control system considering communication quantization and stochastic delay," *Asian J. Control*, vol. 25, no. 5, pp. 3616–3631, Sep. 2023, doi: 10.1002/asjc.3045.
[14] J. Pan, Q. Xu, K. Li, and J. Wang, "Cloud Control of Connected Vehicle Under Both Communication-Induced Sensing and Control Delay: A Prediction Method," *IEEE Trans. Veh. Technol.*, vol. 72, no. 7, pp. 8471–8485, Jul. 2023, doi: 10.1109/TVT.2023.3248870.
[15] Z. Wang, "Motion Estimation of Connected and Automated Vehicles under Communication Delay and Packet Loss of V2X Communications," in *SAE Technical Paper Series*, 400 Commonwealth Drive, Warrendale, PA, United States: SAE International, Apr. 2021. doi: 10.4271/2021-01-0107.
[16] Y. Fang, H. Min, X. Wu, W. Wang, X. Zhao, and G. Mao, "On-Ramp Merging Strategies of Connected and Automated Vehicles Considering Communication Delay," *IEEE Trans. Intell. Transp. Syst.*, vol. 23, no. 9, pp. 15298–15312, Sep. 2022, doi: 10.1109/tits.2022.3140219.
[17] M. Tian, Q. Zhang, D. Tian, L. Jin, J. Li, and F. Xiao, "Pre-Stability Control for In-Wheel-Motor-Driven Electric Vehicles With Dynamic State Prediction," *IEEE Trans. Intell. Veh.*, vol. 9, no. 3, 2024.
[18] D. Hetzer *et al.*, "5G connected and automated driving: use cases, technologies and trials in cross-border environments," *EURASIP J. Wirel. Commun. Netw.*, vol. 2021, no. 1, p. 97, Dec. 2021, doi: 10.1186/s13638-021-01976-6.
[19] K. Sasaki, S. Makido, and A. Nakao, "Vehicle Control System for Cooperative Driving Coordinated Multi-Layered Edge Servers," in *2018 IEEE 7th Int. Conf. on Cloud Netw. (CloudNet)*, Tokyo: IEEE, Oct. 2018, pp. 1–7. doi: 10.1109/CloudNet.2018.8549396.
[20] R. Wang, X. Zhang, Z. Xu, X. Zhao, and X. Li, "Research on Performance and Function Testing of V2X in a Closed Test Field," *J. Adv. Transp.*, vol. 2021, pp. 1–18, Aug. 2021, doi: 10.1155/2021/9970978.
[21] S. Chen, J. Hu, Y. Shi, L. Zhao, and W. Li, "A Vision of C-V2X: Technologies, Field Testing, and Challenges With Chinese Development," *IEEE Internet Things J.*, vol. 7, no. 5, pp. 3872–3881, May 2020, doi: 10.1109/JIOT.2020.2974823.
[22] D. H. Cao, M. S. Gangakhedkar, M. A. Ali, M. M. Gharba, and M. J. Eichinger, "A Testbed for Experimenting 5G-V2X Requiring Ultra Reliability and Low-Latency," in *Proc. IEEE WSA*, Mar. 2017, pp. 1-4.
[23] X. Zhang, R. Wang, X. Zhao, Z. Xu, and F. Zeng, "Experimental Study on Performance of V2X Communication Applied in Typical Traffic Systems in a Closed Test Scenario," in *Proc. CICTP 2020*, Aug. 2020, pp. 812–824. doi: 10.1061/9780784482933.070.
[24] V. Maglogiannis, D. Naudts, S. Hadiwardoyo, D. Van Den Akker, J. Marquez-Barja, and I. Moerman, "Experimental V2X Evaluation for C-V2X and ITS-G5 Technologies in a Real-Life Highway Environment," *IEEE Trans. Netw. Serv. Manag.*, vol. 19, no. 2, pp. 1521–1538, Jun. 2022, doi: 10.1109/TNSM.2021.3129348.
[25] C. Weiß, "V2X communication in Europe – From research projects towards standardization and field testing of vehicle communication technology," *Comput. Netw.*, vol. 55, no. 14, pp. 3103–3119, Oct. 2011, doi: 10.1016/j.comnet.2011.03.016.
[26] B. Di, L. Song, Y. Li, and G. Y. Li, "Non-Orthogonal Multiple Access for High-Reliable and Low-Latency V2X Communications in 5G Systems," *IEEE J. Sel. Areas Commun.*, vol. 35, no. 10, pp. 2383–2397, Oct. 2017, doi: 10.1109/JSAC.2017.2726018.
[27] M. Emara, M. C. Filippou, and D. Sabella, "MEC-Assisted End-to-End Latency Evaluations for C-V2X Communications," in *Proc. 2018 European Conf. on Netw. and Commun. (EuCNC)*, Jun. 2018, pp. 1–9. doi: 10.1109/EuCNC.2018.8442825.
[28] Q. Ye, W. Zhuang, X. Li, and J. Rao, "End-to-End Delay Modeling for Embedded VNF Chains in 5G Core Networks," *IEEE Internet Things J.*, vol. 6, no. 1, pp. 692–704, Feb. 2019.
[29] B. Cucor, P. Kamencay, M. Dado, and T. Petrov, "Experimental Comparison of 4G and 5G Technologies for Connected and Automated Vehicles," in *Proc. 33rd Int. Conf. Radioelektronika (RADIOELEKTRONIKA)*, Apr. 2023, pp. 1–5.
[30] M. C. Lucas-Estan, B. Coll-Perales, T. shimizu, J. Gozalvez, T. Higuchi, S. Avedisov, O. Altintas, and M. Sepulcre, "An Analytical Latency Model and Evaluation of the Capacity of 5G NR to Support V2X Services Using V2N2V Communications," *IEEE Trans. Veh. Technol.*, vol. 72, no. 2, pp. 2293–2306, Feb. 2023.
[31] M. Skocaj, F. Conserva, N. Sarcone, A. Orsi, D. Micheli, G. Ghinamo, S. Bizzarri, and R. Verdone, "Data-driven Predictive Latency for 5G: A Theoretical and Experimental Analysis Using Network Measurements," in *Proc. IEEE Int. Symp. on Personal, Indoor and Mobile Radio Commun. (PIMRC)*, Aug. 31, 2023.
[32] B. Coll-Perales, M. Lucas-Estan, T. Shimizu, J. Gozalvez, T. Higuchi, S. Avedisov, O. Altintas, and M. Sepulcre, "End-to-End V2X Latency Modeling and Analysis in 5G Networks," *IEEE Trans. Veh. Technol.*, vol. 72, no. 4, pp. 5094–5109, Apr. 2023, doi: 10.1109/TVT.2022.3224614.
[33] V. Todisco, S. Bartoletti, C. Campolo, A. Molinaro, A. O. Berthet, and A. Bazzi, "Performance Analysis of Sidelink 5G-V2X Mode 2 Through an Open-Source Simulator," *IEEE Access*, vol. 9, pp. 145648–145661, 2021, doi: 10.1109/ACCESS.2021.3121151.
[34] D. Raca, D. Leahy, C. J. Sreenan, and J. J. Quinlan, "Beyond throughput, the next generation: a 5G dataset with channel and context metrics," in *Proc. of the 11th ACM Multimedia Syst. Conf.*, May 2020.
[35] J. Santa *et al.*, "Evaluation platform for 5G vehicular communications," *Veh. Commun.*, vol. 38, p. 100537, Dec. 2022.
[36] F. Glust, G. Verin, K. Antevski, *et al*, "MEC Deployments in 4G and Evolution Towards 5G," FR, 979-10-92620-18-4, Feb. 2018.
[37] M. H. Almannaa, H. chen, H. A. Rakha, A. Loulizi, I. El-Shawarby, "Field implementation and testing of an automated eco-cooperative adaptive cruise control system in the vicinity of signalized intersections," *Transport Res D-TR E.*, vol. 67, pp244-262, Feb. 2019.
[38] Z. Xu, X. Li, X. Zhao, M. H. Zhang, and Z. Wang, "DSRC versus 4G-LTE for Connected Vehicle Applications: A Study on Field Experiments of Vehicular Communication Performance," *J. Adv. Transp.*, vol. 2017, pp. 1–10, 2017, doi: 10.1155/2017/2750452.
[39] A. Kousaridas *et al.*, "QoS Prediction for 5G Connected and Automated Driving," *IEEE Commun. Mag.*, vol. 59, no. 9, pp. 58–64, Sep. 2021, doi: 10.1109/MCOM.110.2100042.